\documentstyle[pre,twocolumn,aps,epsf]{revtex} 

\font\tenmsb=msbm10
\def\Bbb#1{\hbox{\tenmsb#1}}

\begin{document}
\draft

\title{Mechanisms of synchronization and pattern formation in
a lattice of pulse-coupled oscillators}

\author{Albert D\'{\i}az-Guilera\cite{albertmail}, Conrad J.
P\'{e}rez\cite{conradmail} }
\address{
Departament de F\'{\i}sica Fonamental, 
Universitat de Barcelona, 
Diagonal 647, E-08028 Barcelona, Spain \\}
\author{Alex Arenas\cite{alexmail} }
\address{
Departament d'Enginyeria Inform\`{a}tica, Universitat
Rovira i Virgili, Carretera Salou s/n, E-43006  Tarragona, Spain \\}
\date{\today}
\maketitle

\begin{abstract}
We analyze the physical mechanisms leading either to synchronization
or to the formation of spatio-temporal patterns in a lattice model of
pulse-coupled oscillators. In order to make the system tractable from
a mathematical point of view we study a one-dimensional ring with
unidirectional coupling. In such a situation, exact results
concerning the stability of the fixed of the dynamic evolution of the
lattice can be obtained.
Furthermore, 
we show that this stability is the responsible for the different
behaviors.
%We study analytically the dynamic behavior of a population of
%pulse-coupled oscillators defined in a one-dimensional ring. We put
%special attention in the physical mechanisms leading either to
%synchronization or to spatio-temporal structures. We also analyze
%the
%stability of these dynamical regimes remarking the relevance of the
%coupling in the long time features of the system.
\end{abstract}


\section{Introduction}

Among the collective phenomena that are currently attracting the
interest of the scientific community one of the most relevant
concerns the synchronization of the temporal activity of
populations of interacting nonlinear oscillators, due to its
ubiquity in many different fields of science. Experimental
evidences of this phenomenon have been reported for centuries
\cite{Huy} but in the last decades the advance in the
comprehension of its nature has allowed to develop a
theoretical description. In this context, several successful
ideas has been suggested. An interesting approach proposed in
\cite{Win,Kura84,Erm} has been shown to
be useful to describe the dynamic evolution of the population.
The idea consist in modeling the system as an assembly of phase
oscillators interacting through continuous-time couplings. For
sufficiently large coupling strength the system may undergo a
phase transition from incoherence to spontaneous mutual
synchronization. More challenging from a theoretical and
realistic point of view is to consider networks of pulse-coupled
oscillators which may account for the behavior of heart
pacemaker cells, integrate and fire neurons, and other systems
made of excitable units. The intrinsic nonlinearities associated
to these models make their dynamical evolution more difficult to
describe and only in the last years real advances have occurred
\cite{Pes,Mir,Kura91,Abb,Tre}.

Up to now, almost all the theoretical approaches have been centered
around
mean-field models or populations of just a few oscillators. From
these
studies it is possible to investigate the mechanisms relevant 
for the formation
of assemblies of synchronized elements as well as other
spatio-temporal
structures. However, these mean-field descriptions are, in many
cases, far
from reality and other methods where the specific topology or
geometry of
the system, as well as the precise connectivity between units, must 
be
considered because their effects may be crucial. In a such new world
many
points remain open. In particular the majority of works rely on
simulations showing the outstanding richness that a low dimensional
system
of pulse-coupled oscillators may display. Some examples are
self-organized
criticality, chaos, quasiperiodicity, etc. \cite{rev}.

Unfortunately, a rigorous mathematical description of these systems
is still
missing. Some of the theoretical papers appeared in the scientific
literature proof the stability of some behaviors \cite{Gold,PhysD}
but they do not
explore the
mechanisms leading to them. The goal of this paper goes in both
directions:
the analysis of the mechanisms which are responsible for
synchronization and
formation of spatio-temporal structures, and, as a complement, to
proof
under which conditions they are stable solutions of the dynamical
equations.
Since our motivation is to analyze the essence of the problem we have
considered a 1D model which will allow us to illustrate the ideas in
a very
clear way.
In spite of this apparent simplicity,
this system displays a rich set of behaviors, that depend on the
specific values of
the
parameters of the model, which has been observed in lattices with
higher coordination numbers.\cite{PhysD,PRL2}
Notice that populations of 1D pulse coupled
oscillators are currently of great interest in some areas of science.
As an
example let us mention that for a certain type of cardiac arrhythmia
there
is an abnormally rapid heartbeat whose period is set by the time that
an
excitation takes to travel a circuit. This observation can be
explained by
modeling appropriately the circulation of a wave of excitation on a
one-dimensional ring \cite{Glass}. In a different context,
synchronization and periodic states of 1D populations of phase-locked
loops have been recently investigated.\cite{Gold,Vieira} 

The structure of this paper is as follows. In Sec. II we describe the
system as well as the notation used throughout the paper. Sections
III and
IV are devoted to analyze the simplest cases of three and four
oscillators, respectively. In Sec. V we study the general case,
whereas in the last section we present our conclusions.

\section{The model}

Let us consider a system formed by a population of $(N+1)$
oscillators
distributed on a ring. The state of each oscillator is described by
its
phase which increases linearly in time, until one of them reaches a
threshold value that, without loss of generality, we have considered
equal
to $\phi _{th}=1$. When this happens the oscillator fires and changes
the
state of its rightmost neighbor according to  
\[
\phi _{i}\geq 1\Rightarrow \left\{ 
\begin{array}{l}
\phi _{i}\rightarrow 0 \\ 
\phi _{i+1}\rightarrow \phi _{i+1}+\varepsilon \phi _{i+1}\equiv \mu
\phi
_{i+1}
\end{array}
\right. 
\]
\begin{equation}
\hspace{5em}\forall i=0,\ldots ,N
\end{equation}
subjected to periodic boundary conditions, i.e. $N+1\equiv 0$,
and where $\varepsilon $ denotes the
strength of the coupling. From an effective point of view, the
pulse-interaction between oscillators, as well as the state of each
unit of
the system, can be described in terms of changes in the phase, or in
other
words, in terms of the so called phase response curve (PRC), $\mu
\phi$ in
our case. Behind this fact one assumes that the phase shift elicited
by an
impulse affects the period of a given unit in the current time
interval
but not in future intervals. In this paper we have also considered a
linear PRC \cite{PRC}.
In practise, however, this condition can be relaxed
since
a nonlinear PRC does not change the qualitative behavior of the model
provided the number of fixed points of the dynamics is not altered.
Moreover, a linear PRC has the advantage of making the system
tractable
from an analytical point of view. 

Let us describe the notation used in the paper. The population is
ordered
according to the following criterion: The oscillator which fires will
be
always labeled as unit 0 and the rest of the population will be
ordered from
this unit clockwise. After the firing, the system is driven until
another
oscillator reaches the threshold. Then, we relabel the units such
that the
oscillator at $\phi =1$ is now unit number 0, and so on. The whole
process
can be described through a suitable transformation. This fact will
enable us
to study the origin of different structures in a very simple way. Our
strategy has been to trace the phases of the oscillators after each
firing
and then to construct return maps either of a complete cycle, in
which all
the oscillators fire exactly once, or after a single firing + driving
process (FD). Let us clarify this point mathematically. The first
step is
to construct the matrix of the transformation for a FD. To illustrate
the 
situation let us consider the general transformation for a 'jump' $n$
between two successive firings, distinguishing between $n=1$

\vspace{5mm}

\begin{tabular}{llllll}
$\phi_0=1$ & $\rightarrow$ & $0$ & $\rightarrow$ & $1-\mu\phi_{1}$ &
$=
\phi^{\prime}_{N}$ \\ 
$\phi_1$ & $\rightarrow$ & $\mu\phi_1$ & $\rightarrow$ & $1$ &
$= \phi^{\prime}_{0}$  \\ 
$\phi_2$ & $\rightarrow$ & $\phi_2$ & $\rightarrow$ &
$1-\mu\phi_1+\phi_2$ & 
$=\phi^{\prime}_1$ \\ 
$\vdots$ &  & $\vdots$ &  & $\vdots$ & $\vdots$ \\ 
$\phi_i$ & $\rightarrow$ & $\phi_i$ & $\rightarrow$ &
$1-\mu\phi_{1}+\phi_i$
& $= \phi^{\prime}_{i-1}$ \\ 
$\vdots$ &  & $\vdots$ &  & $\vdots$ & $\vdots$ \\ 
$\phi_N$ & $\rightarrow$ & $\phi_N$ & $\rightarrow$ &
$1-\mu\phi_{1}+\phi_N$
& $= \phi^{\prime}_{N-1}$%
\end{tabular}

\vspace{5mm}

\noindent and $n>1$

\vspace{5mm}

\begin{tabular}{llllll}
$\phi_0=1$ & $\rightarrow$ & $0$ & $\rightarrow$ & $1-\phi_{n}$ & $=
\phi^{\prime}_{N-n+1}$ \\ 
$\phi_1$ & $\rightarrow$ & $\mu\phi_1$ & $\rightarrow$ & $%
1-\phi_{n}+\mu\phi_1$ & $= \phi^{\prime}_{N-n+2}$ \\ 
$\phi_2$ & $\rightarrow$ & $\phi_2$ & $\rightarrow$ &
$1-\phi_n+\phi_2$ & $%
=\phi^{\prime}_{N-n+3}$ \\ 
$\vdots$ &  & $\vdots$ &  & $\vdots$ & $\vdots$ \\ 
$\phi_i$ & $\rightarrow$ & $\phi_i$ & $\rightarrow$ &
$1-\phi_{n}+\phi_i$ & $%
= \phi^{\prime}_{N-n+i+1}$ \\ 
$\vdots$ &  & $\vdots$ &  & $\vdots$ & $\vdots$ \\ 
$\phi_{n-1}$ & $\rightarrow$ & $\phi_{n-1}$ & $\rightarrow$ & $%
1-\phi_{n}+\phi_{n-1}$ & $= \phi^{\prime}_{N}$ \\ 
$\phi_{n}$ & $\rightarrow$ & $\phi_{n}$ & $\rightarrow$ & $1$ & =
$\phi^{\prime}_0$  \\  
$\phi_{n+1}$ & $\rightarrow$ & $\phi_{n+1}$ & $\rightarrow$ & $%
1-\phi_{n}+\phi_{n+1}$ & $= \phi^{\prime}_{1}$ \\ 
$\vdots$ &  & $\vdots$ &  & $\vdots$ & $\vdots$ \\ 
$\phi_j$ & $\rightarrow$ & $\phi_j$ & $\rightarrow$ &
$1-\phi_{n}+\phi_j$ & $%
= \phi^{\prime}_{j-n}$ \\ 
$\vdots$ &  & $\vdots$ &  & $\vdots$ & $\vdots$ \\ 
$\phi_N$ & $\rightarrow$ & $\phi_N$ & $\rightarrow$ &
$1-\phi_{n}+\phi_N$ & $%
= \phi^{\prime}_{N-n}$%
\end{tabular}
\vspace{5mm} \newline
where $\phi^{\prime}$ describe the new phases after the FD process.
The
diagram describes the situation just when the leader fires (first
column),
the change in phases as a consequence of the emitted pulse (second)
and
finally the evolution of the system due to the linear driving up to
the
next
firing (third). We believe that this is the simplest and most compact
way
to depict the process since we get rid of rotations that should be
taken into account after the linear driving for any other relabeling
method. Thus the transformation that describes this process reads

\[
{\vec{\phi}}^{\prime}= T_n(\vec{\phi}) \equiv \vec{1}+{\Bbb{M}}_n
\vec{\phi},
\]
where $\vec{\phi}^{\prime}$ is a vector with $N$ components since the
zero-th
component does not play any role in the description. In the above
expression ${\Bbb{M}}_n$ is an $N\times N$ matrix that can be written
as 

\begin{equation}
\left( {\Bbb{M}}_1 \right)_{ij} =  \delta _{i+1,j} - (1+\varepsilon)
\delta_{j,1}
\label{defm1}
\end{equation}

\begin{equation}
\left( {\Bbb{M}}_n \right)_{ij} = \delta_{i+n,j} - \delta_{j,n}
+\varepsilon \delta_{j,1}\delta_{i+n,1} \hspace{2em} \forall n>1.
\label{defmn}
\end{equation}
In these expressions $\delta _{i,j}$ is the usual Kronecker delta.
The sums
should be interpreted modulus (N+1) and none of the subscripts can be
either 
$0$ or $N+1$. 

Since we are interested in emphasizing the mechanisms leading either
to
synchronization or pattern formation, we have considered very
convenient to
start our discussion with two illustrative situations where everything
can be
computed analytically and whose perfect understanding will help us to
tackle
the general case.

\section{Three oscillators}

This is the simplest case which is worth analyzing, since the system
formed by two units has been widely analyzed in the literature, see
for instance \cite{Mir,PhysD,Gras}. If we define a simple cycle as a
sequence of firings in which each oscillator fires once and only
once, there are only two possibilities for this system:
\begin{itemize}
\item  A: 0,1,2

\item  B: 0,2,1
\end{itemize}

\noindent provided oscillator 0 is always at the threshold at the
starting
point of the dynamic evolution. Here, the numbering corresponds to
the
firing sequence according to the
initial spatial order in the lattice. Let us study both situations in
detail.

\subsection{Order A: 0,1,2}

In this case, the sequence starts when oscillator 0 fires, sending a
pulse
that changes the state of oscillator 1. Afterwards, the system is
driven
until oscillator 1 arrives to the threshold. According to our
notation
this process can be viewed as:
\begin{eqnarray}
1  \rightarrow & 0 & \rightarrow 1-\mu\phi_1=\phi'_2  \nonumber \\
\phi_1  \rightarrow & \mu\phi_1 & \rightarrow 1  \nonumber \\
\phi_2  \rightarrow & \phi_2 & \rightarrow
1+\phi_2-\mu\phi_1=\phi'_1.
\end{eqnarray}

Then, we have transformed a state characterized by two phases
$\phi_1$ and $%
\phi_2$ to a new one also characterized by two phases
$\phi^{\prime}_1$ and $%
\phi^{\prime}_2$ such that $\phi^{\prime}_1$ is always the phase of
the
oscillator that will receive the next pulse and keep this numeric
order
along the ring. In matrix notation the transformation can be written
as
follows

\begin{equation}
\left( 
\begin{array}{c}
\phi^{\prime}_1 \\ 
\phi^{\prime}_2
\end{array}
\right) = \left( 
\begin{array}{c}
1 \\ 
1
\end{array}
\right) + \underbrace{\left( 
\begin{array}{cc}
-\mu & 1 \\ 
-\mu & 0
\end{array}
\right) }_{{\Bbb{M}}_1} \left( 
\begin{array}{c}
\phi_1 \\ 
\phi_2
\end{array}
\right).
\end{equation}
The complete cycle is constructed by applying three times this
transformation [$T_1\circ T_1\circ T_1 (\phi)$]. In other words 
\begin{equation}
\vec{\phi}^{\prime\prime\prime}=\vec{R}_A+{\Bbb{M}}_A \cdot
\vec{\phi}.
\end{equation}
The independent term $\vec{R}_A$ is 
\begin{equation}
\vec{R}_A = \left( {\vec{\vec{1}}} + {\Bbb{M}}_1 + {\Bbb{M}}_1 \cdot 
{\Bbb{M}}_1 \right) \cdot \vec{1},
\end{equation}
where $\vec{1}$ is a column vector of 1's, ${\vec{\vec{1}}}$ is the
identity
matrix, and the matrix of the transformation ${\Bbb{M}}_A$ is defined
as 
\begin{equation}
{\Bbb{M}}_A={\Bbb{M}}_1 \cdot {\Bbb{M}}_1 \cdot {\Bbb{M}}_1.
\end{equation}
From this expression it is easy to compute the fixed points of the
transformation, which are solutions of the equation 
\begin{equation}
\vec{\phi}^*=\vec{R}_A+{\Bbb{M}}_A \cdot \vec{\phi}^*,
\end{equation}
that is 
\begin{equation}
\phi^*_1=\frac{2}{3+2 \varepsilon }
\end{equation}
\begin{equation}
\phi^*_2=\frac{1}{3+2 \varepsilon }.
\end{equation}
The stability of these fixed points is given by the eigenvalues of 
${\Bbb{M}}_A$ which are 
\begin{equation}
\frac{-(\mu-3)\mu^2 \pm i(\mu-1)\mu^{3/2}\sqrt{4-\mu}}{2}
\end{equation}
whose moduli are $\mu^{3/2}$. Depending on the sign of $\varepsilon $
the
fixed point is either stable (-) or unstable (+). 

\subsection{Order B: 0,2,1}

Now, oscillator 1 receives the pulse but it is oscillator 2 the one
that
leads the driving and arrives first to the threshold

\begin{eqnarray}
1 \rightarrow & 0 & \rightarrow 1-\phi_2  \nonumber \\
\phi_1  \rightarrow & \mu\phi_1 & \rightarrow 1+\mu\phi_1-\phi_2
\nonumber \\
\phi_2  \rightarrow & \phi_2 & \rightarrow 1
\end{eqnarray}
Therefore, the new phases are

\begin{equation}
\left( 
\begin{array}{c}
\phi _{1}^{\prime } \\ 
\phi _{2}^{\prime }
\end{array}
\right) =\left( 
\begin{array}{c}
1 \\ 
1
\end{array}
\right) +\underbrace{\left( 
\begin{array}{cc}
0 & -1 \\ 
\mu & -1
\end{array}
\right) }_{\Bbb{M}_{2}}
\left( 
\begin{array}{c}
\phi _{1} \\ 
\phi _{2}
\end{array}
\right)
\end{equation}

Again the complete cycle is constructed by applying three times this
transformation [$T_2\circ T_2\circ T_2 (\phi)$]: 
\begin{equation}
\vec{\phi}^{\prime \prime \prime }=\vec{R}_{B}+{\Bbb{M}}_{B}\cdot
\vec{\phi}
\end{equation}
where the independent term $\vec{R}_{B}$ now is 
\begin{equation}
\vec{R}_{B}=\left( \vec{\vec{1}}+{\Bbb{M}}_{2}+{\Bbb{M}}_{2}\cdot 
{\Bbb{M}}_{2}\right) 
\cdot \vec{1}
\end{equation}
and the matrix of the transformation 
\begin{equation}
{\Bbb{M}}_{B}={\Bbb{M}}_{2}\cdot {\Bbb{M}}_{2}\cdot {\Bbb{M}}_{2}.
\end{equation}
The fixed point of this transformation, which is solution of the
equation 
\begin{equation}
\vec{\phi}^{*}=\vec{R}_{B}+{\Bbb{M}}_{B}\cdot \vec{\phi}^{*},
\end{equation}
is
\begin{equation}
\phi _{1}^{*}=\frac{1}{3+\varepsilon }
\end{equation}
\begin{equation}
\phi _{2}^{*}=\frac{2+\varepsilon }{3+\varepsilon }.
\end{equation}
The stability of this fixed points is given by the eigenvalues of 
${\Bbb{M}}_{B}$ that are 
\begin{equation}
\frac{3\mu -1\pm i(\mu -1)\sqrt{4\mu -1}}{2}
\end{equation}
whose moduli are again $\mu ^{3/2}$. Therefore, both fixed points
describe
the same physical behavior that is independent on the particular
order in
which oscillators fire.

\subsection{Phase ordering}

In the previous subsections we have studied some features of the
final
state of the system when a sequence of identical transformations are
applied successively. However the reader can argue that those
sequences are
not the only possible dynamical evolution. Indeed, to deal with all
the
possible situations we should analyze what happens when a mixture of
$T_1$
and $T_2$ are combined in an arbitrary manner to complete a cycle,
and
what sort of physical consequences derive from this fact. In
addition, one
may wonder whether an advancement can take place, i.e. if a given
oscillator can fire twice before another element of the chain arrives
to
the threshold, breaking thus our definition of a simple cycle.
Such issues are discussed in this subsection. To
illustrate
this point let us start by considering Fig. 1, where we have plot the
evolution of the phases each time oscillator 0
is at the threshold value. It is obvious that there are
two
different situations: a positive or negative value of the coupling.
In this figure we have analyzed the situation for $\varepsilon < 0$.

On the left hand side of Fig.~1 we have all the possible initial
configurations and
on its
right hand side how they do transform when oscillator 0 is again at
the
threshold. Regions A and B
represent the sequences A and B described before, respectively. Here,
we
can see that states lying initially in one of these regions will 
approach the fixed points (attractors), since A and B are slowly
shrinking. Therefore, once one starts with a given sequence no other
one
can be applied to describe the dynamical evolution of the system. The
physical picture associated to the attractor fixed point is  quite
simple.
The oscillators remain at a certain distance in the phase space
(phase-locking). For
larger dimensions (more oscillators) this fact induces the creation
of
complex spatial patterns.

\begin{figure}[h]
\vspace{8cm}
\epsfxsize=3.5truein\epsffile{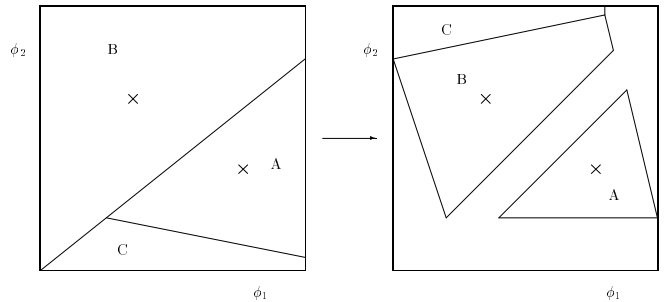}
\caption{Evolution of the phases $\phi_1$ and $\phi_2$, for a
negative value of
$\varepsilon$. Right:
initial configurations when oscillator 0 is at the threshold value;
left: new values of the phases when 
oscillator 0 reaches the threshold again.
The crosses correspond to the fixed points and each
region corresponds to a given sequence of firings (see text).} 
\end{figure}

A special attention deserves region C which corresponds to a sequence
of
firings 010. This means that oscillator 0 advances oscillator 2, a
situation not covered before. Here we can see that the effect
of this advancement is to reorganise the phases in such a way that
after
one cycle the old configurations fall in the basin of attraction of
region
B and therefore, for them, sequence B must be applied forever. No
more
advancements can take place. The main conclusion is that advancements
play
a role only in the transient but not in the stationary properties of
the
dynamics. A more clear picture of the physical meaning of this fact
will
be provided in the next section.  

The mechanism for positive $\varepsilon$ is the opposite that for
negative 
$\varepsilon$. Regions A and B are enlarged every cycle, and the
configurations move away the fixed points (repellers) until they
cross
some of the borders where at least two oscillators get absorbed and
synchronize. Fig.~2 shows this fact. The left hand side of the figure
depicts the size of the basin of attraction of those configurations
(closed lines) that when oscillator 0 reaches the threshold again
still require sequence A
or B
to evolve dynamically. The rest of the phase space is formed by
states
characterized by the fact that after the next firing of oscillator 0
occurs at least two units
will merge.
These units are specified (underlined) on the right hand side of the
figure. Since two synchronized oscillators act as a single one and
cannot be broken after a complete cycle, the problem now is
equivalent to
that of two oscillators. This dimensional reduction is the essence of
synchronization. For mean-field models the word absorption has been
coined
to illustrate this phenomenon \cite{Mir}.

\begin{figure}[tbh]
%\vspace{8cm}
\epsfxsize=3.5truein\epsffile{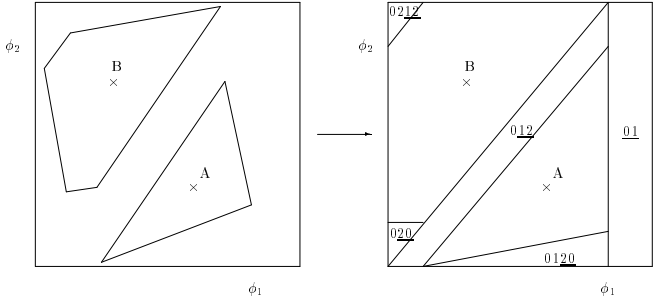}
\caption{The same as Fig. 1 for positive coupling.}
\end{figure}

To address the question about the plausibility of having mixtures of
consecutive transformations $T_1$ and $T_2$ it is convenient
to look at the problem from another perspective. Instead of
considering
complete cycles it is better to analyze single firings. Let us
suppose for
simplicity negative (inhibitory) coupling between units. In this
case, it
is evident that the alternative application of both transformations
lead
to two possible options. The first case to be considered is the
combination $T_1 \circ T_2$. We can observe that applying $T_2$ is
inconsistent with the application of
$T_1$ afterwards; because the resulting configuration
$T_2(\vec{\phi})$ does not satisfy the possible different phase
orders necessary to apply $T_1$. On the other hand, the order $T_2
\circ T_1$ implies one advancement between oscillators, corresponding
to region C (see Fig.~1) of phase space.  This situation has been
discussed previously to play the role of a transient dynamic behavior
of the system.
Thus we can conclude that, in general, the advancements will make the
phases to be reordered until the system reaches a configuration which
is consistent with only one sequence of transformations.

\section{Four oscillators}

In this case there are 6 different orders for the oscillators to
complete a simple cycle:

\begin{itemize}
\item  A: 0,1,2,3

\item  B: 0,1,3,2

\item  C: 0,2,1,3

\item  D: 0,2,3,1

\item  E: 0,3,1,2

\item  F: 0,3,2,1
\end{itemize}

Analogously to the 3 oscillators case we define the matrices of the
transformation between successive firings of two oscillators,
according to
the jump between oscillators that fire successively:

\begin{equation}
{\Bbb{M}}_1=\left( 
\begin{array}{ccc}
-\mu & 1 & 0 \\ 
-\mu & 0 & 1 \\ 
-\mu & 0 & 0
\end{array}
\right)
\end{equation}

\begin{equation}
{\Bbb{M}}_2=\left( 
\begin{array}{ccc}
0 & -1 & 1 \\ 
0 & -1 & 0 \\ 
\mu & -1 & 0
\end{array}
\right)
\end{equation}

\begin{equation}
{\Bbb{M}}_3=\left( 
\begin{array}{ccc}
0 & 0 & -1 \\ 
\mu & 0 & -1 \\ 
0 & 1 & -1
\end{array}
\right)
\end{equation}

We can easily compute the eigenvalues of those matrices. For
${\Bbb{M}}_{1}$
and ${\Bbb{M}}_{3}$ the eigenvalues have moduli larger (smaller) than
1 for
positive $\varepsilon $ (negative $\varepsilon $). However, for
${\Bbb{M}}_{2}$ there is one
eigenvalue with modulus equal to 1.
This will be very important when discussing the
stability of the fixed points.

According to these jumps the transformation of a complete cycle for
the
different orders are constructed in the following ways

\begin{itemize}
\item  A: 0,1,2,3 $\rightarrow $ $T_1 \circ T_1 \circ T_1 \circ T_1$

\item  B: 0,1,3,2 $\rightarrow $ $T_2 \circ T_{3} \circ T_{2} \circ
T_{1}$

\item  C: 0,2,1,3 $\rightarrow $ $T_{1} \circ T_{2} \circ T_{3} \circ
T_{2}$

\item  D: 0,2,3,1 $\rightarrow $ $T_{3} \circ T_{2} \circ T_{1} \circ
T_{2}$

\item  E: 0,3,1,2 $\rightarrow $ $T_{2} \circ T_{1} \circ T_{2} \circ
T_{3}$

\item  F: 0,3,2,1 $\rightarrow $ $T_{3} \circ T_{3} \circ T_{3} \circ
T_{3}$
\end{itemize}

For instance, let us write, in matrix form, the transformation for
the case
B:
\[
\vec{\phi}^{\prime}= \vec{1} + {\Bbb{M}}_2 \cdot \vec{1} + {\Bbb{M}}%
_2 \cdot {\Bbb{M}}_3 \cdot \vec{1} +{\Bbb{M}}_2 \cdot {\Bbb{M}}%
_3 \cdot {\Bbb{M}}_2 \cdot \vec{1} +
\]
\begin{equation}
\hspace{5em}+ {\Bbb{M}}_2 \cdot {\Bbb{M}}%
_3 \cdot {\Bbb{M}}_2 \cdot {\Bbb{M}}_1 \cdot \vec{\phi}.
\end{equation}

We can now proceed to compute the fixed points associated to these
transformations which are:

\begin{itemize}
\item  A: $1$, $\frac{3}{4+3\varepsilon }$, $\frac{2}{4+3\varepsilon
}$, $%
\frac{1}{4+3\varepsilon }$.

\item  B: $1$, $\frac{1}{2+\varepsilon }$, $0$, $\frac{1+\varepsilon
}{%
2+\varepsilon }$.

\item  C: $1$, $\frac{1}{2+\varepsilon }$, $1$,
$\frac{1}{2+\varepsilon }$.

\item  D: $1$, $\frac{1}{2+\varepsilon }$, $1$,
$\frac{1}{2+\varepsilon }$.

\item  E: $1$, $\frac{1}{2+\varepsilon }$, $0$, $\frac{1+\varepsilon
}{%
2+\varepsilon }$.

\item  F: $1$, $\frac{1}{4+\varepsilon }$, $\frac{2+\varepsilon }{%
4+\varepsilon }$, $\frac{3+\varepsilon }{4+\varepsilon }$.
\end{itemize}

Numerically, these fixed points are unique since this is ensured by
the fact that all the eigenvalues of the matrix that multiplies
$\vec{\phi}$ are different from 1 in all the cases. Nevertheless the
fixed points for cases C and D are senseless since they do no verify
the prescribed order, and hence they are physically unacceptable.
Furthermore, the fixed points of cases B and C are physically the
same since they just differ in the order in which oscillators 0 and 2
fire but they are synchronized. Thus, basically, we have to deal with
fixed points where the phase difference between adjacent oscillators
is roughly $\phi_{i+1}-\phi_i=1/4,2/4,3/4 (\mbox{mod } 1)$, with
small corrections depending on $\varepsilon $.

Let us focus now on the process from the point of view of single
firings again. Notice that $T_{1}$ and $T_{3}$ are transformations
that can be applied alone successively leading to a natural complete
cycle. However, $T_{2}$ can not be applied in this way because this
would lead to an unphysical situation where only a group of
oscillators will fire.
This means that $T_{2}$ must be combined with $T_{1}$ or $T_{3}$ in
order to be physically acceptable, then the effect of having an
eigenvalue with modulus 1, that can carry a metastability on the
system, is avoided by this combination. The combinations of $T_{2}$
with either $T_{1}$ or $T_{3}$, can be done in the way described by
the orders B, C ,D or E, that always give rise to a chessboard type
pattern. Other combinations can be obtained applying $T_{1}\circ
T_{2}$ or $T_{3}\circ T_{2}$ forever, this situation elicits
advancements but nevertheless the resulting patterns are chessboard
type
again.

\section{N+1 oscillators}

Although, in general, to construct a complete cycle is not a trivial
mechanism we can infer some keys about the behavior of the system
from the
FD processes as we have done for the simplest lattices: compute the
fixed points and their stability.

\subsection{Fixed points} 

Due to their different behavior we will have to
distinguish again
the cases $n=1$ and $n>1$. Thus, for the first situation, we have to
solve

\begin{equation}
\left. 
\begin{array}{l}
\phi_1=1-\mu\phi_1+\phi_2 \\ 
\phi_2=1-\mu\phi_1+\phi_3 \\ 
\hspace{1cm}\vdots \\ 
\phi_{N-1}=1-\mu\phi_1+\phi_N \\ 
\phi_N=1-\mu\phi_1
\end{array}
\right\}
\end{equation}
Simply summing up all the equations we are left with
$\phi_1=N(1-\mu\phi_1)$
and then we get

\begin{equation}
\phi_1^*=\frac{N}{N+1+N\varepsilon} .
\label{fp1}
\end{equation}

Notice that the other phases at the fixed point can be obtained from
this
one, since $\phi_{j-1}-\phi_{j}=1-\mu\phi_1 \;\forall \;
j=2,\ldots,N$. This fixed point corresponds to a situation in which
all the oscillators fire in turn following their lattice ordering.

On the other hand, for $n>1$ we have to solve

\begin{equation}
\left. 
\begin{array}{l}
\phi_1=1-\phi_n+\phi_{n+1} \\ 
\hspace{1cm}\vdots \\ 
\phi_{N-n}=1-\phi_n+\phi_N \\ 
\phi_{N-n+1}=1-\phi_n \\ 
\phi_{N-n+2}=1-\phi_n+\mu\phi_1 \\ 
\phi_{N-n+3}=1-\phi_n+\phi_2 \\ 
\hspace{1cm}\vdots \\ 
\phi_N=1-\phi_n+\phi_{n-1}
\end{array}
\right\}
\end{equation}

Summing up again all the equations we now obtain

\begin{equation}
\phi_1+\phi_n=N(1-\phi_n)+\mu\phi_1.
\end{equation}

Now we notice that it is not enough to get one of the values of the
phases.
We have to close the system of equation by means of the following
procedure:

\[
\phi_1=1-\phi_n+\underbrace{\phi_{n+1}}_{1-\phi_n+\phi_{2n+1}}=
2(1-\phi_n)+\phi_{2n+1} 
\]
and so on. Again all the subscripts are understood modulus $N+1$.
This
procedure is repeated until we reach $\phi_{N-n+1}$ which closes the
dependence of $\phi_1$ on $\phi_n$. Obviously, a necessary condition
to
close it is that $N+1$ and $n$ do not have common factors. This
procedure is iterated $m_n - 1$ times, where $m_n$ verifies
\[
(1+m_nn) \mbox{ mod } (N+1)=0, 
\]
and it exists and is unique for each $n<N+1$. Then for a given $N$ we
will
have
to consider all the values of $m_n$ between $1$ and $N$ without
common
factors. We can therefore obtain that at the fixed point

\begin{equation}
\phi_n^*=\frac{N+m_n \varepsilon}{N+1+m_n \varepsilon } \hspace{2em}
\mbox{and} 
\hspace{2em} \phi_1^*=\frac{m_n}{N+1+m_n \varepsilon}.
\label{fpn}
\end{equation}

Again, this situation would correspond to a sequence of FD processes
of jump $n$. It is easy to convince oneself that in both cases ($n=1$
and $n>1$) one can build a complete cycle by applying $N+1$ times the
transformation $T_n$ which does not change the fixed points. In
principle, this successive application could make new fixed points to
appear, but this is forbidden by the fact that the
moduli of the eigenvalues of $\Bbb{M}_n$ is always larger than 1 for
$\varepsilon > 0$, and smaller than 1 in the opposite case, whenever
$N+1$ and $n$ do not
have common factors, as we show in the appendix.

But we still do not know what happens when $N+1$ and $n$ have common
factors. As we show in the appendix in this case there exists at
least one eigenvalue of modulus 1 and this fact does not ensure us
about the existence of a solution for the set of algebraic equations;
even when this solution can exist it usually gives rise to unphysical
situations, as, for instance, to values of the phases that are either
below zero or above one. We showed explicitely for the case of 4
oscillators that under these circumstances the transformation
of an FD process of this kind has to be combined with other
transformations with noncommon factors and that this leaded to
advancement between oscillators and to the formation of a chessboard
like pattern. This is also what happens in the general case. The
matrix associated to this combination will have eigenvalues with
moduli different from 1 and will guarantee the existence of the fixed
point. Let us assume that a given spatial structure with period
$p=(N+1)/n$ exists and then there are some oscillators which are
synchronized
($\phi_0=\phi_p=\phi_{2p}=\ldots=\phi_{(n-1)p};\;\phi_1=\phi_{p+1}=
\ldots;\;\ldots$). Hence we need only to consider each spatial
period, since the transformations of jump $p$ correspond to
oscillators that fire exactly at the same time. Within each period
there will be different possibilities for the magnitude of the jumps
$n_p$ and then the fixed points will be characterized by
\begin{equation}
\phi_1^*=\frac{m_{n_p}}{p+m_{n_p}\varepsilon}=\frac{nm_{n_p}}{N+1+nm_
{n_p} \varepsilon}.
\label{fpp}
\end{equation}

By combining Eqs. (\ref{fp1}), (\ref{fpn}), and
(\ref{fpp}), we realize that for a given $N+1$ there always
exists a value of $m$ ($0<m<N+1$) such that
\begin{equation}
\phi_1^*=\frac{m}{N+1+m\varepsilon} 
\label{fpgen}
\end{equation}
will be the phase of oscillator 1 at a fixed point. We have used this
fact to identify the fixed points of the dynamics in simulations of
lattices of a few oscillators, as we will see later on. 

In principle one can still think on the possibility of other fixed
points corresponding to combinations of transformations not described
above. For instance, successive applications of two transformations of
different values of $n$; but this case will necessarily involve
advancements between the oscillators, which, as we have already
discussed, are only important in the transient but not in the
approach to the final state. This, of course, can make the transients
to become quite large, as we have observed in the computer
simulations, but the only final states for an inhibitory coupling are
those described earlier in the text. It is also important to note
that depending on the strength of the coupling, and on the number of
oscillators, there will be some fixed points, for an inhibitory
coupling, that will not exist; those that verify
\begin{equation}
\varepsilon<1-\frac{N+1}{m}.
\end{equation}
For instance, for the three oscillators case this happens for $m=2$
when $\varepsilon < -0.5$ [see (10)] and then region A disappears and
the fixed point corresponding to B is the only possible final state.

\subsection{Stability of the fixed points}

After having shown the existence of the fixed points for single FD
processes and extending this calculation to complete cycles, one
needs to compute their stability. Since the calculation of the
eigenvalues of the matrices is a lengthy but straightforward
procedure we have left it for an appendix. There we show that the
results easily obtained for three and four oscillators also apply to
the general case, i.e. for $\varepsilon <0$ the fixed points are
attractors, whereas in the opposite case they are repellers. On the
one hand, the attractiveness of the fixed points enables the
formation of spatio-temporal patterns of phase-locked oscillators. 
This works not only for the structures with different phases but also
for the periodic ones. 
On the other hand, when these fixed points become repellers it makes
that neighboring oscillators synchronize and from that time on they
will act as a single unit; this absorption (or dimensional reduction
in our language) is iterated until the whole system acts as a single
unit which completes the mechanism of the synchronization of
the lattice models with very-short range interactions we have
analyzed through the paper.

\subsection{Computer simulations}

In order to check the validity of our results we have made computer
simulations on lattices of a few oscillators. In Tables I and II we
represent the percentage of the structures the system formed by $N+1$
oscillators reaches as a stationary state for two different values of
$\varepsilon$ as a functions of $m$ which stands roughly for the
phase difference between neighboring oscillators times $N+1$, see
(\ref{fpgen}). There are several results in these simulations that
deserve further comments. For instance we can notice that the
oscillators tend to keep the maximum phase difference. The chessboard
like structure has the largest basin of attraction when the
population has an even number of oscillators; whereas for an odd
number of oscillators there are two peaks with the largest phase
differences. However these results depend slightly on the strength of
the coupling since the maximum percentage appears for the maximum
phase difference and the larger the peak the larger the phase
difference. Thus we can understand the different behavior for the two
different values of $\varepsilon$. Another difference concerns the
reduction and, eventually, the disappearance of the basins of
attraction of the fixed points that correspond to large values of
$m$. This fact also affects the time the system needs to reach the
stationary state; for instance, for smaller values of $\varepsilon$
not only the jumps along one cycle are smaller but there are also
more attractive fixed points. On the other hand, we have corroborated
that for excitatory couplings the only possible final state is
synchronization, no matter how long the transient is.

\begin{table}[b]
\caption[]{Percentage of the final states the system formed by $N+1$
oscillators reaches, for $\varepsilon$=-0.1. The first column stands
for $N+1$ and the first row for $m$ in (\ref{fpgen}), which
approximately corresponds to the phase difference times $N+1$ between
consecutive oscillators. It is averaged for 1000 initial random
configurations picking each phase from a uniform ditribution between
0 and 1.}
\begin{tabular}{ccccccccccc}
& 1 & 2 & 3 & 4 & 5 & 6 & 7 & 8 & 9 & 10 \\ \hline
3 & 61.4 & 38.6 &  &  &  &  &  &  &  &  \\ 
4 & 21.2 & 71.3 & 7.5 &  &  &  &  &  &  &  \\ 
5 & 5.5 & 56.5 & 37.2 & 0.8 &  &  &  &  &  &  \\ 
6 & 1.2 & 29.1 & 58.0 & 11.7 & 0.0 &  &  &  &  &  \\ 
7 & 0.0 & 11.3 & 51.7 & 34.2 & 2.5 & 0.0 &  &  &  &  \\ 
8 & 0.0 & 3.8 & 32.1 & 50.6 & 13.2 & 0.4 & 0.0 &  &  &  \\ 
9 & 0.0 & 1.1 & 16.2 & 47.9 & 30.8 & 3.9 & 0.1 & 0.0 &  &  \\ 
10 & 0.0 & 0.2 & 6.4 & 34.6 & 43.8 & 14.3 & 0.7 & 0.0 & 0.0 & \\
12 & 0.0 & 0.0 & 0.7 & 9.2 & 34.7 & 40.0 & 14.0 & 1.4 & 0.0 & 0.0 \\
15 & 0.0 & 0.0 & 0.0 & 0.5 & 5.9 & 23.0 & 38.8 & 25.2 & 6.2 & 0.4 \\
\end{tabular}
\end{table}

\begin{table}[b]
\caption{The same as Table I for $\varepsilon$=-0.01.}
\begin{tabular}{cccccccccc}
& 1 & 2 & 3 & 4 & 5 & 6 & 7 & 8 & 9  \\ \hline
3 & 51.9 & 48.1 &  &  &  &  &  &  &  \\ 
4 & 17.5 & 67.3 & 15.1 &  &  &  &  &  &  \\
5 & 4.7 & 46.7 & 45.0 & 3.61 &  &  &  &  &  \\ 
6 & 1.0 & 22.8 & 54.8 & 20.7 & 0.7 &  &  &  &  \\ 
7 & 0.2 & 8.4 & 43.0 & 40.9 & 7.4 & 0.1 &  &  &  \\ 
8 & 0.0 & 2.8 & 24.3 & 48.3 & 22.5 & 2.06 & 0.0 &  &  \\ 
9 & 0.0 & 0.8 & 11.35 & 39.7 & 37.7 & 9.7 & 0.6 & 0.0 &  \\ 
10 & 0.0 & 0.2 & 4.4 & 26.0 & 42.9 & 23.1 & 3.4 & 0.1 & 0.0 \\
\end{tabular}
\end{table}

\section{Conclusions}

In order to analyze the mechanisms of synchronization and the
formation of spatio-temporal structures we have introduced a very
simple model of pulse-coupled oscillators: a one-dimensional ring
with unidirectional coupling. Despite this apparent simplicity it
conserves all the features of low-dimensional systems subjected to
short-range interactions which develop large-scale structures.

Although the dynamic evolution of the system involves two time
scales, a slow one for the driving and a fast one for the
interaction, we have constructed return maps that gives a complete
information of the system. Concerning the maps, we have been able to
compute exactly the fixed points of the dynamical evolution and their
stability.

For a negative (inhibitory) coupling the fixed points are attractors
of the dynamics. Each one of these attractors has a well defined
basin of attraction, although in some cases those regions are not
simply connected. Since the evolution is discrete there are jumps
among non-connected regions that correspond to advancements between
oscillators. The advancements are only important in the transient
dynamics, until the phases of the oscillators lie in the final basin
of attraction of the fixed point. The volumes of the basins of
attraction are different and depend on the value of the
coupling, as we have checked by means of computer simulations of a
few oscillators. However, the states with the maximum phase
difference between neighboring oscillators seems to be the preferred
ones. 

On the other hand, for a positive (excitatory) coupling the fixed
points are repellers of the dynamical evolution. Although in a
configuration space with a multiplicity of repellers one can think
that the system will jump from one region to another this is not our
case. There are absorbing barriers surrounding the repellers; when
the system reaches one of these barriers it means that at least two
neighboring oscillators have synchronized. When this happens the set
of synchronized oscillators acts as a single unit that cannot be
broken. From that time on we only need to consider a reduced number
of units; we call this fact dimensional reduction since the new
system can be described in terms of matrices with less components.
This process of absorption is iterated until the system reaches a
completely synchronized configuration.

The present work only concerns the qualitative behavior of a
population of pulse-coupled oscillators; nevertheless, a quantitative
behavior about the time a given population needs to reach the
stationary state, either a synchronized one or a spatio-temporal
pattern, would be desirable in order to complete the description of
such systems. Another interesting question is related to the
stability of the different structures with respect to fluctuations.

\acknowledgments
This work has been supported by
CICyT of the Spanish Government, grant \#PB94-0897.

\appendix
\section*{}

In this appendix we will compute the bounds of the eigenvalues of the
matrices ${{\Bbb{M}}_n}$ defined in Eqs. (2)-(3). All the
information will be extracted form the characteristic polynomial
$P_{N,n}( \lambda ,\varepsilon)$ which corresponds to the determinant
$| \lambda \vec{\vec{1}} - {{\Bbb{M}}_n} |$. In this case we will
distinguish several situations. First of all when $N+1$ and $n$ have
common factors it is easy to see in the determinant that always
exists a minor that corresponds to eigenvalues of modulus 1; anyway,
as we explain in the text we do not need to care about this case. 

The case $n=1$ is very simple to compute. By simple inspection of the
determinant it is easy to see that
\begin{equation}
P_{N,1}( \lambda ,\varepsilon)= \lambda P_{N-1,1}( \lambda
,\varepsilon)+\mu
\end{equation}
and hence
\begin{equation}
P_{N,1}( \lambda ,\varepsilon)= \lambda ^N+\mu  \lambda ^{N-1} + \mu
\lambda ^{N-2} +
\ldots + \mu  \lambda  +\mu.
\label{pn1}
\end{equation}

In order to compute the case $n>1$ it is convenient to introduce the
following similarity transformation matrix

\[
(B_n)_{i,j}=\delta_{j,m_n i},
\]
where again the subscripts are understood modulus $N+1$,
such that when it is applied to $M_n$, $M'_n=B^{\dagger}
\cdot M_n\cdot B$, one gets

\[
(M'_n)_{i,j}=\delta_{i,j+1}-\delta_{j,N}+\varepsilon\delta_{j,m_n}
\delta_{i,m_n+1}.
\]
With this procedure we have converted a, in principle, complex matrix
in a much simpler one. The new matrix has $-1$ in the last column and
$1$'s just below the main diagonal, unless at column $m_n$ and row
$m_n+1$ where it has $1+\varepsilon$. In order to compute the
eigenvalues we should notice that when $\varepsilon=0$ the
characteristic polynomial is
\[
P_{N,n}( \lambda ,\varepsilon=0)= \lambda ^N+ \lambda ^{N-1} + 
\lambda ^{N-2} +
\ldots +  \lambda  + 1.
\]
Then to compute it for $\varepsilon\ne 0$ we expand the determinant
around the previous case and one realizes that the columns that are
at the right and the rows that are below this element will not
contribute. Thus one has
\[
P_{N,n}( \lambda ,\varepsilon)= \lambda ^N+ \lambda ^{N-1} + 
\lambda ^{N-2} +
\ldots +  \lambda ^{m_n}  +
\]
\begin{equation}
\hspace{5em}+\mu ( \lambda ^{m_n-1} + \ldots + 1).
\label{pnn}
\end{equation}
The last expression can be taken as general, i.e. including the $n=1$
case,  bearing in mind that $m_1=N$.

It is obvious that to compute the roots of (\ref{pnn}) is an
unnecessary task, since the only needed information concerning the
stability are the bounds of
the eigenvalues. In order to compute these bounds we will look at the
properties of the characteristic polynomial. It can be rewritten as

\begin{eqnarray}
P_{N,n}( \lambda ,\varepsilon)&=& \sum_{k=0}^{N} \lambda ^k +
\varepsilon \sum_{k=0}^{m_n-1} \lambda ^k=\nonumber\\
&=&
\frac{ \lambda ^{N+1}-1}{ \lambda -1}+\varepsilon\frac{ \lambda
^{m_n}-1}{ \lambda -1}.
\end{eqnarray}
Then the eigenvalues of the matrix will correspond to the
roots of
\[
( \lambda ^{N+1}-1)+\varepsilon( \lambda ^{m_n}-1)=0
\]
unless the artificially introduced $ \lambda =1$. These roots will
verify
\[
| \lambda ^{N+1}+\varepsilon \lambda ^{m_n}|=1+\varepsilon.
\]
This modulus has upper and lower bounds given by the sum and the
difference, i.e.
\[
|\lambda|^{N+1}-|\varepsilon||\lambda|^{m_n}\le 1+\varepsilon
\le
|\lambda|^{N+1}+|\varepsilon||\lambda|^{m_n}.
\]
Thus for $\varepsilon >0$, we use
\[
|\lambda|^{N+1}+\varepsilon|\lambda|^{m_n}\ge 1+\varepsilon
\]
which implies that $|\lambda|\ge 1$. On the other hand, for
$\varepsilon <0$ we take
\[
|\lambda|^{N+1}-|\varepsilon||\lambda|^{m_n}\le 1-|\varepsilon|
\]
which, in turn, implies that $|\lambda|\le 1$.
The final point is to show that the equality can only be fulfilled
when both complex numbers have the same direction. It is easy to see
that this can only happen when $N+1$ and $m_n$ have common factors.
Since this fact is avoided in this demostration we are left with the
fact that for an excitatory coupling ($\varepsilon>0$) the
eigenvalues are larger than 1 and the opposite for an inhibitory
coupling ($\varepsilon<0$), as we wanted to show.


\begin{references}
\bibitem{PRC} This assumption has physical sense in some situations.
For instance, for the Peskin model of pacemaker cardiac 
cells \cite{Pes} one can expand the PRC \cite{PhysD} in powers
of the convexity of the driving, leading to a PRC with a constant 
term, whose only effect is to shift the threshold, plus another term
which is proportional to the phase.
\end{references}
\end{document}